\documentclass[11pt,english]{article}

\newcommand{\CO}{{\cal O}}

\newcommand{\CL}{{\cal L}}

\makeatletter
\newcommand*{\rom}[1]{\expandafter\@slowromancap\romannumeral #1@}
\makeatother

\usepackage[latin9]{inputenc}
\usepackage{geometry}
\usepackage{verbatim}
\usepackage{prettyref}
\usepackage[numbers,sort&compress]{natbib}
\usepackage{amsmath}
\usepackage{amssymb}
\usepackage{flexisym}
\usepackage{cases}
\usepackage{ytableau}
\usepackage{graphicx}
\usepackage{caption}
\usepackage{subcaption}
\usepackage{tikz}
 \usepackage{slashed}
\usepackage{esvect}
\usepackage{dsfont}
\usepackage{color}
\definecolor{darkgreen}{rgb}{0,0.5,0}
\definecolor{darkblue}{rgb}{0,0,0.6}
\definecolor{purple}{rgb}{0.4,.2,0.7}
\usepackage[colorlinks=true,citecolor=blue,linkcolor=blue,urlcolor=blue]{hyperref}

\usepackage{tocloft}
\hypersetup{
  linktoc=none % prevents coloring of TOC entries
}
 	
\numberwithin{equation}{section}
\numberwithin{figure}{section}
\numberwithin{table}{section}

\def\CQ{{\cal Q}}

\DeclareFontShape{OT1}{cmr}{mx}{n}{<->cmr10}{}

\begin{document}

\fontseries{mx}\selectfont

\begin{center}

\,

\,

\,

\LARGE \bf The phase of charged Nariai solutions

\end{center}

\vskip2cm

\begin{center}
Victor Ivo$^{1}$ and  Zimo Sun$^{1,2}$
\vskip5mm
{\it{\footnotesize $^{1}$ Jadwin Hall, Princeton University, Princeton, NJ 08540, USA} \\
\it{\footnotesize $^2$ Institute for Advanced Study, Princeton, NJ 08540, USA}\\
}
\end{center}

\vskip2cm

\begin{abstract}
In this note, we compute the phase of the one-loop Euclidean path integral around charged Nariai solutions in 4 dimensions, including both metric and gauge field fluctuations. These solutions have a $S^{2} \times S^{2}$ geometry, and a magnetic flux in one of the spheres. For charges smaller than a critical value, the phase matches the result for the uncharged Nariai solution, and for charges bigger than that value, the phase is $i^{3}$. Our analytical calculation in the full 4D geometry matches the result obtained recently within a 2D dilaton gravity reduction. Along the way, we also develop a method of dealing with residue zero modes in the de Donder gauge.

\end{abstract}

\newpage

\tableofcontents

\section{Introduction}

Understanding quantum gravity in de Sitter spacetime is a very important problem, but without a UV-complete theory of de Sitter gravity, it is unclear which principles one should hold sacred. One might hope, however, that in the absence of such a UV-complete realization, we can still learn about de Sitter using semiclassical Einstein gravity. After all, the recent developments in the black hole information paradox emphasize that the gravitational path integral \cite{Almheiri:2019qdq, Penington:2019kki} is a valuable tool.

The oldest development in this direction was the proposal by Gibbons and Hawking \cite{Gibbons:1976ue} that the Euclidean path integral can be interpreted as a partition function of a microscopic theory describing gravity. An immediate obstruction to this proposal, however, is that Euclidean gravity has a well-known conformal factor problem \cite{Gibbons:1978ac}. This implies that to regularize the path integral, one needs some contour rotation prescription, which generally implies the path integral has a nonvanishing phase \cite{Hawking:2010nzr, Polchinski:1988ua}.

Since this phase is present in the one-loop Euclidean path integral around the sphere saddle \cite{Polchinski:1988ua}, $Z(S^{D})$, it is unclear if one can interpret $Z(S^{D})$ as a partition function of de Sitter. Motivated by this puzzle, Maldacena proposed including a probe observer in the Euclidean sphere saddle to cancel part of the phase \cite{Maldacena:2024spf}\footnote{Recently, it was proposed that the observer might actually cancel the entire phase, see \cite{Chen:2025jqm}.}. This proposal led to a new wave of interest in the phase of Euclidean gravity partition functions \cite{Ivo:2025yek, Shi:2025amq, Anninos:2025ltd, Law:2025yec, Ivo:2025fwe, Chen:2025jqm}.

\begin{figure}
    \centering
    \includegraphics[width=0.9\linewidth]{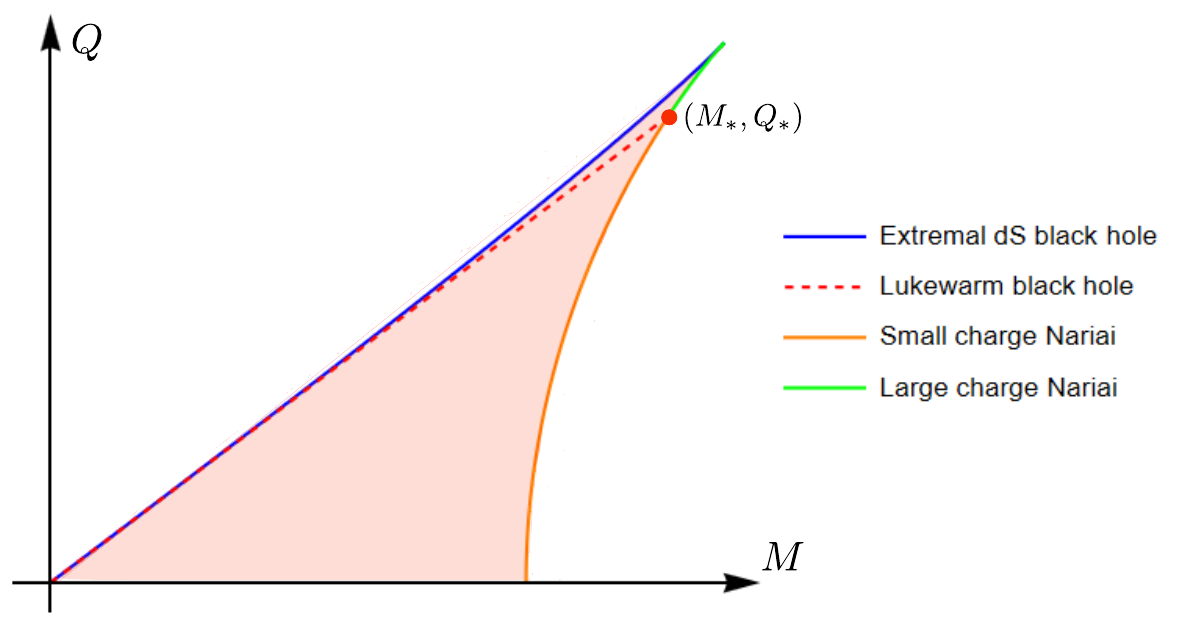}
    \caption{We plot the allowed values of the mass and charge $(M, Q)$ for black holes in de Sitter for a given value of the cosmological constant $\Lambda>0$. The black hole solutions are points in the shaded region delimited by the blue and orange $+$ green curves. The blue curve corresponds to extremal black holes, that is, the black holes with the smallest value of mass for a given charge. The dashed red line stands for the so-called ``lukewarm" solutions, which are black holes in thermal equilibrium with the cosmological horizon. The orange and green curves correspond to the ``Nariai solutions", which are $dS_{2}\times S^{2}$ geometries. The orange and green colours stand for whether their charge is smaller or larger than a threshold value of $Q=Q_{*}=\frac{1}{4}\sqrt{\frac{3}{\Lambda}}$, respectively. The transition between the curves is marked with a red dot.}
    \label{skfin}
\end{figure}

In a recent paper by Chen, Stanford, Tang, and Yang \cite{Chen:2025jqm}, the authors analyzed the phase of the Euclidean path integral around charged black hole solutions in de Sitter. The analysis was done in a dilaton gravity reduction with the U(1) field being on-shell. To be more specific, charged black holes that can be analytically continued to smooth geometries in the  Euclidean signature have a relation between charge and mass defined by the lukewarm and Nariai solutions delimited in the ``Shark Fin" in Figure \ref{skfin}. 

In this note, we perform the exact analytic calculation of the phase for 4$D$ solutions in one of the branches they studied, namely the charged Nariai branch. This solution has a $dS_{2}\times S^{2}$ geometry, and a magnetic flux in the $S^{2}$. We study its Euclidean counterpart where the $dS_{2}$ factor is analytically continued to $S^{2}$.

Our strategy for computing the phase will be to follow the prescription in \cite{Ivo:2025yek} (or equivalently \cite{Shi:2025amq})\footnote{This is the same prescription used by the authors of \cite{Chen:2025jqm}.}. This boils down to counting negative modes. Being more explicit, each negative mode contributes to the phase with a factor of $(-i)$, so the phase would be $(-i)^{n_{E}}$ with $n_{E}$ the number of negative modes. Since $n_{E}$ is infinite, we subtract from it an infinite local counterterm, $n_{c}$, which is roughly the ``number of points in spacetime"\footnote{This idea goes back to Polchinski \cite{Polchinski:1988ua}. A regularized version of this procedure was implemented rigorously in \cite{Chen:2025jqm}.}. This implies that the system can have a phase equivalent to a ``negative number" of negative modes.

We will work in a specific gauge fixing scheme where the calculation is especially simple. In this special gauge, we will have some zero modes unfixed by the gauge condition. By perturbing the gauge condition slightly, we will lift all zero modes and obtain the correct phase, confirming the result of \cite{Chen:2025jqm} for solutions along the Nariai branch. 

The paper is structured as follows. In section \ref{classgem} we introduce the charged Nariai solution. Then, in section \ref{quadacsec} we discuss the quadratic action around this solution, and the specific gauge fixing scheme that we will be using. In particular, we highlight the presence of 6 ghost zero modes for our specific gauge choice.  In section \ref{modexp} we simplify the quadratic action by doing an explicit mode expansion of the relevant fields. After doing so, the quadratic action becomes especially simple in terms of the mode expansion coefficients. We will diagonalize the action somehow explicitly and highlight the presence of 6 bosonic zero modes specific to our gauge choice. Finally, in section \ref{defdeDon} we will deal with the residual zero modes by perturbing the gauge condition slightly, and thus find the final answer for the overall phase. In section \ref{disc}, we conclude with some final remarks.

\section{The classical geometry}
\label{classgem}
The Euclidean Einstein-Maxwell action in $4$ dimensions reads:
\begin{align}\label{EMaction}
S = - \frac{1}{16\pi G}\int d^4 x\sqrt{g}\left(\mathcal R-2\Lambda- F_{\mu\nu}F^{\mu\nu}\right)~.
\end{align}
This action has diffeomorphism invariance and the U(1) gauge symmetry. The diffeomorphism acts as a Lie derivative on both the metric and the U(1) gauge field:
\begin{align}\label{g1}
\CL_\xi g_{\mu\nu} =\nabla_\mu\xi_\nu + \nabla_\nu \xi_\mu , \quad \CL_\xi A_\mu = - \xi^\nu F_{\mu\nu}+\partial_\mu(\xi\cdot A),
\end{align}
where the total derivative term $\partial_\mu(\xi\cdot A)$ can be absorbed into a U(1) gauge transformation.

We are interested in the magnetically charged Nariai saddle of \eqref{EMaction}. The geometry is a product of two $S^2$, described by the following metric
\begin{align}
ds^2 = L^2 \left(d\tau^2+\sin^2\tau d\phi^2\right)+R^2 \left(d\theta^2+\sin^2\theta d\varphi^2\right)~.
\end{align}
Take the U(1) gauge field background to be $F = Q \sin\theta d\theta \wedge d\varphi$, i.e. $F_{\theta\varphi} = Q\sin\theta$.
The second sphere is thus magnetically charged. The radii $L$ and $R$ are fixed by the equation of motion 
\begin{align}\label{EOM1}
{\cal R}_{\mu\nu}- \frac{1}{2}g_{\mu\nu} {\cal R}+\Lambda g_{\mu\nu}= 2F_{\mu\rho}F_\nu^{\,\,\rho} - \frac{1}{2} F^2 g_{\mu\nu}~,
\end{align}
where $F^2=F_{\mu\nu}F^{\mu\nu}=\frac{2Q^2}{R^4}$.
The right-hand side of \eqref{EOM1} is traceless because Maxwell theory is scaling invariant in 4$D$. The tracelessness yields $\mathcal R=4\Lambda$, which allows us to rewrite \eqref{EOM1} as
\begin{align}\label{EOM2}
{\cal R}_{\mu\nu}- \Lambda g_{\mu\nu} =2F_{\mu\rho}F_\nu^{\,\,\rho} - \frac{1}{2} F^2 g_{\mu\nu} = \frac{Q^2}{R^4}\begin{pmatrix}-g_{\alpha\beta} & 0\\ 0 & g_{\alpha'\beta'}\end{pmatrix}~.
\end{align}
where $\alpha, \beta$ correspond to coordinates on the first sphere and $\alpha', \beta'$ correspond to coordinates on the second sphere. Using \eqref{EOM2}, we obtain 
\begin{align}\label{LR}
\Lambda-\frac{Q^2}{R^4} = \frac{1}{L^2}, \quad \Lambda+\frac{Q^2}{R^4} = \frac{1}{R^2}~.
\end{align}
 When $Q=0$, both radii are equal to $\frac{1}{\sqrt{\Lambda}}$. By turning on a magnetic charge on the second sphere, the second sphere shrinks while the first sphere grows. 
The explicit solutions  of \eqref{LR} are 
\begin{align}\label{LRsolu}
L^2=\frac{1+\sqrt{1-4\Lambda Q^2}}{2\Lambda\sqrt{1-4\Lambda Q^2}}, \quad R^2=\frac{1+\sqrt{1- 4\Lambda Q^2}}{2\Lambda}~.
\end{align}
The maximum magnetic charge is $Q_{\rm max}=\frac{1}{2\sqrt{\Lambda}}$. As $Q$ approaches $Q_{\rm max}$, the first sphere blows up and the second sphere reaches the minimal radius $\frac{1}{\sqrt{2\Lambda}}$. In this limit, the geometry becomes $\mathbb R^2\times S^2$.

The on-shell action of a charged Nariai solution is 
\begin{align}
    S_{\rm on-shell} = -\frac{2\pi R^2}{G} = -\frac{\pi(1+\sqrt{1-4\Lambda Q^2})}{\Lambda G}~.
\end{align}
$-S_{\rm on-shell}$ is bounded from above by $\frac{2\pi}{\Lambda G}$, which corresponds to the $Q=0$ limit. Even though the first sphere blows up as $Q$ approaches $Q_{\rm max}$, the on-shell stays finite in this limit.

\section{The quadratic action in the de Donder gauge}
\label{quadacsec}
To perform the one-loop path integral, we first need to derive the quadratic action around the Nariai background. 
Expand the metric and $U(1)$ gauge field around their saddle point configuration:
\begin{align}
g_{\mu\nu} \to  g_{\mu\nu} +\sqrt{32\pi G}\, \phi_{\mu\nu}, \quad A_\mu \to   A_\mu + \sqrt{4\pi G}a_\mu.
\end{align}
Plugging these expansions into \eqref{EMaction}, we
find  the  full quadratic action 
\begin{align}\label{S2f}
S^{(2)}  = \int d^4 x \sqrt{g}&\left[\frac{1}{2}\phi^{\mu\nu}\Delta_2 \phi_{\mu\nu}\!+\!\frac{1}{2}\phi(\nabla^2+\Lambda)\phi\! -\! \nabla^\nu\phi_{\mu\nu}\nabla_\rho\phi^{\mu\rho}-\!\phi\nabla_\mu\nabla_\nu\phi^{\mu\nu}\!+\!\phi_{\mu\nu}(R^{\mu\rho}\phi_{\rho}^{\,\,\nu}\!-\!\Lambda\phi^{\mu\nu})\right.\nonumber\\
&\left.+\frac{F^2}{2}\left(\phi^{\mu\nu}\phi_{\mu\nu}-\frac{1}{2}\phi^2\right)+2F_{\mu\nu}F_{\rho\sigma}\phi^{\mu\rho}\phi^{\nu\sigma}+\frac{1}{4} f^2+\frac{\phi F^{\mu\nu} f_{\mu\nu}- 4 f_{\mu\nu} F_\rho^{\,\,\nu} \phi^{\mu\rho}}{\sqrt{2}}\right]~,
\end{align}
where $\Delta_{2}\phi_{\mu\nu} = -\nabla^2 \phi_{\mu\nu} -2 R_{\mu\rho\nu\sigma} \phi^{\rho\sigma}$. This quadratic action reduces to its flat space counterpart \cite{Monteiro:2008wr} in the $\Lambda\to 0$ limit, after making the field redefinition: $\phi_{\mu\nu} \to \frac{1}{\sqrt{2}} \phi_{\mu\nu}$ and $a_\mu \to 2 a_\mu$.
At the quadratic level, the gauge symmetry of the original Einstein-Maxwell action becomes
\begin{align}\label{g2}
{\rm Diff}: \,\,\begin{cases}\delta_\xi \phi_{\mu\nu} = \frac{1}{2}\left(\nabla_\mu\xi_\nu + \nabla_\nu \xi_\mu\right)\\ \delta_\xi a_\mu = -\sqrt{2}\, \xi^\nu F_{\mu\nu}
\end{cases}, \quad {\rm U(1)}:\,\, \begin{cases}\delta_\chi \phi_{\mu\nu} =0\\ \delta_\chi a_\mu = \partial_\mu \chi
\end{cases}~.
\end{align}
For example, one can remove the U(1) fluctuations on the second sphere by using part of the Diff symmetry. However, this is not how we fix the gauge symmetry. In this section, we will add the standard de Donder gauge fixing term $(\nabla^\nu\phi_{\mu\nu} -\frac{1}{2}\nabla_\mu\phi)^2$ for the diffeomorphism, because it gives a particularly simple action.
We will add $\frac{1}{2}(\nabla^\mu a_\mu)^2$ to fix the U(1) symmetry. After a lengthy calculation, we find the gauge-fixed action to be \footnote{Our convention for the antisymmetric tensors on the two spheres is $\epsilon_{\tau\phi} = L^2\sin\tau, \epsilon_{\theta\varphi} = R^2\sin\theta$. }
\begin{align}\label{SdD2}
S^{(2)}_{\rm dD} &= \int \frac{1}{2}\psi^{\alpha\beta}\left(-\nabla^2\!+\!\frac{2}{L^2}\right)\psi_{\alpha\beta}+ \frac{1}{2}\psi^{\alpha'\beta'}\left(-\nabla^2\!+\!\frac{2}{R^2}\right)\psi_{\alpha'\beta'}
+\frac{1}{2}a^\mu(-\nabla^2 a_\mu \!+\! R_{\mu\nu} a^\nu)+\frac{1}{8}\phi(\nabla^2\!+\!2\Lambda)\phi\nonumber\\
&+ \psi^{\alpha\beta'}\left(-\nabla^2\!+\!F^2\right)\psi_{\alpha\beta'}-2\sigma(\nabla^2\!+\!2\Lambda\!-\!2F^2)\sigma-F^2\phi\sigma+\frac{2\sqrt{2}Q}{R^2}\epsilon^{\alpha'\beta'}\left(f_{\alpha\alpha'} \psi^\alpha_{\,\,\beta'} +\sigma f_{\alpha'\beta'}\right)~.
\end{align}
where we have used the following decomposition for the metric fluctuation $\phi_{\mu\nu}$:
\begin{align}
\phi_{\mu\nu} = \begin{pmatrix}\psi_{\alpha\beta} +(\phi/4+\sigma)g_{\alpha\beta} & \psi_{\alpha\beta'} \\ \psi_{\alpha'\beta} & \psi_{\alpha'\beta'} +(\phi/4-\sigma)g_{\alpha'\beta'}   \end{pmatrix}~.
\end{align}
$\psi_{\alpha\beta}$ is traceless on the first sphere and $\psi_{\alpha'\beta'}$ is traceless on the second sphere. $\phi$ is the trace of $\phi_{\mu\nu}$ and $\sigma$ captures the fluctuation of the relative size of the two spheres.

Let us also comment on the ghost action. To write down the ghost action, we use the BRST formalism. The action of the BRST charge $\CQ_{\rm BRST}$ is given by 
\begin{align}
    \CQ_{\rm BRST}\, (\phi_{\mu\nu}) = \frac{1}{2}\left(\nabla_\mu c_\nu + \nabla_\nu c_\mu\right), \quad \CQ_{\rm BRST}\,( a_\mu) = \nabla_\mu c - \sqrt{2} F_{\mu\nu} c^\nu~,
\end{align}
where $c_\mu$ is the ghost field associated with the Diff symmetry, and $c$ is the U(1) ghost. Let $\bar c_\mu$ and $\bar c$ be the corresponding anti-ghost fields, which are annihilated by the BRST charge. The ghost Lagrangian is thus 
\begin{align}
    \CL_{\rm ghost} &= \CQ_{\rm BRST}\left(\bar c^\mu\left(\nabla^\nu \phi_{\mu\nu} -\frac{1}{2}\nabla_\mu\phi\right)+\bar c \nabla_\mu a^\mu\right)\nonumber\\
    &=\bar c^\mu (-\nabla^2 c_\mu - R_{\mu\nu} c^\nu)+\bar c(-\nabla^2 c)+\sqrt{2}\,\bar c \,\nabla_\mu c_\nu F^{\mu\nu}~.
\end{align}
The ghost action has some trivial zero modes corresponding to the Killing vectors and constant U(1) transformations, which lead to the division of gauge group volumes. There are additional zero modes with $c=0$ and $c_\mu$  longitudinal, i.e. $c_\mu = \nabla_\mu \eta$. The scalar function $\eta$ should satisfy $
\nabla_\mu \nabla^2\eta + 2 R_{\mu\nu}\nabla^\nu \eta = 0$. 
Decomposing this equation into the two $S^2$ yields 
\begin{align}
\nabla_\alpha\left(\nabla^2 + \frac{2}{L^2}\right)\eta = 0, \quad \nabla_{\alpha'}\left(\nabla^2 + \frac{2}{R^2}\right)\eta = 0~.
\end{align}
The solutions are $(\ell_1, \ell_2)=(1, 0)$ and $(\ell_1, \ell_2) = (0, 1)$, where $\ell_1$ is the angular momentum of $\eta$ on the first sphere and $\ell_2$ is the angular momentum on the second sphere. The first solution is annihilated by $\nabla^2 + \frac{2}{L^2}$ and $\nabla_{\alpha'}$, and the second solution is annihilated by $\nabla^2 + \frac{2}{R^2}$ and $\nabla_{\alpha}$. We will identify the corresponding bosonic zero modes in the next section. As argued by Polchinski in  \cite{Polchinski:1988ua}, the gauge invariance of the partition function requires using the absolute value of the ghost determinant. Therefore, the ghost fields do not contribute to the overall phase of the one-loop path integral.

\section{Mode expansions and zero modes}
\label{modexp}

To compute the phase of the one-loop path integral, we need to identify the negative modes in the quadratic action. It is easy to see the positivity of components like  $\psi_{\alpha\beta}$ and $\psi_{\alpha'\beta'}$, because they are decoupled and their action is manifestly positive. They do not contribute to the phase.
However, for the rest of the components, which mix with each other, it is a more complicated task to find the negative modes.
In this section, we will introduce a mode expansion for these components, allowing us to rewrite the quadratic action as a regular Gaussian integral.
We will then perform diagonalization to disentangle all the modes.

 Let $I=(\ell_1, \ell_2)$ be a collective label, and $\psi_I(x)$ be a scalar mode that has angular momentum $\ell_1$ on the first sphere and $\ell_2$ on the second, i.e.
\begin{align}
&-\nabla_\alpha\nabla^\alpha \psi_I = \lambda_{I, 1}\psi_I, \quad \lambda_{I, 1}=\frac{\ell_1(\ell_1+1)}{L^2}\nonumber\\
& -\nabla_{\alpha'}\nabla^{\alpha'} \psi_I =  \lambda_{I, 2}\psi_I,\quad  \lambda_{I, 2} = \frac{\ell_2(\ell_2+1)}{R^2}~.
\end{align} 
The magnetic quantum number for a given angular momentum is suppressed in this notation.
The normalization of $\psi_I$ is chosen to be  $\int \psi_I \psi_J = \delta_{IJ}$. Define $\lambda_I = \lambda_{I, 1}+\lambda_{I, 2}$.

The scalar components $\phi$ and $\sigma$ can be expanded as follows
\begin{align}\label{scalardec}
\phi(x) = \sum_{I}\phi_I \psi_I (x) ,\quad \sigma(x) = \sum_{I}\sigma_I \psi_I (x) ~.
\end{align}
The vector component, say $a_\alpha$, can be either longitudinal or transverse on the first sphere. For the longitudinal part, the basis is $\nabla_\alpha\psi_I$, and for the transverse part, the basis is $\epsilon_{\alpha\beta}\nabla^\beta \psi_I$.
Therefore, the mode expansion of $a_\mu$ is 
\begin{align}\label{adecom}
a_\alpha = \sum_{I, \ell_1\ge 1} a_I \nabla_\alpha \psi_I +b_I \epsilon_{\alpha\beta}\nabla^\beta \psi_I, \quad a_{\alpha'} = \sum_{I, \ell_2\ge 1} \bar a_I \nabla_{\alpha'} \psi_I +\bar b_I \epsilon_{\alpha'\beta'}\nabla^{\beta'} \psi_I ~. 
\end{align}
The analysis of the off-diagonal components of the metric is similar, but  there are four types of tensor structures depending on whether the mode is transverse or longitudinal on the two spheres. We choose the basis to be
\begin{align}\label{UVWZ}
&{\rm TT}: \,\, U^I_{\alpha\alpha'} = \epsilon_{\alpha\beta} \epsilon_{\alpha'\beta'}\nabla^\beta\nabla^{\beta'}\psi_I, \quad {\rm TL}: \,\, V^I_{\alpha\alpha'} = \epsilon_{\alpha\beta}\nabla^\beta\nabla_{\alpha'}\psi_I,\nonumber\\
& {\rm LT}: \,\, W^I_{\alpha\alpha'} =  \epsilon_{\alpha'\beta'}\nabla_\alpha\nabla^{\beta'}\psi_I, \qquad\,\,\, {\rm LL}: \,\, Z^I_{\alpha\alpha'} = \nabla_\alpha\nabla_{\alpha'}\psi_I~,
\end{align}
and then decompose $\psi_{\alpha\alpha'}$ into the basis
\begin{align}\label{offdec}
\psi_{\alpha\alpha'} = \sum_{I\ge 1} u_I U^I_{\alpha\alpha'} +v_I V^I_{\alpha\alpha'} + w_I W^I_{\alpha\alpha'} + z_I Z^I_{\alpha\alpha'} ~.
\end{align}
Here $I\ge 1$ means that both $\ell_1, \ell_2$ are nonzero.
All  modes in \eqref{UVWZ} satisfy 
\begin{align}
-\nabla^2 \mathcal Y^I_{\alpha\alpha'} = (\lambda_I - 2\Lambda) \mathcal Y^I_{\alpha\alpha'}, \quad \mathcal Y\in\{U, V, W, Z\}~,
\end{align}

Plugging \eqref{scalardec}, \eqref{adecom} and \eqref{offdec} into \eqref{SdD2}, we obtain the mode expansion of the  quadratic action  in the de Donder gauge.
We denote by $\tilde S^{(2)}_{\rm dD}$ the action with the modes
 $\psi_{\alpha\beta}$ and $\psi_{\alpha'\beta'}$ excluded. It reads
\begin{align}
\tilde S^{(2)}_{\rm dD} &=\sum_I \frac{\lambda_I}{2} \left[\lambda_{I, 1}\left(a_I^2 + b_I^2\right)+\lambda_{I, 2}\left(\bar a_I^2 +\bar b_I^2\right)\right]-\frac{\lambda_I-2\Lambda}{8}\phi_I^2+2(\lambda_I+2F^2-2\Lambda)\sigma_I^2 - F^2\phi_I \sigma_I\\
&+\lambda_{I,1}\lambda_{I,2}\left[\left(\lambda_I - \frac{2}{L^2}\right) \left(u_I^2+v_I^2+w_I^2+ z_I^2\right)+\frac{2\sqrt{2}Q}{R^2} ((a_I -\bar a_I)w_I +b_I u_I+\bar b_I z_I)\right] +  \frac{4\sqrt{2} Q}{R^2}\lambda_{I,2}\bar b_I\sigma_I\nonumber
\end{align}
where summations of the form $\sum_I(\cdots)$ are taken over all indices $I$, for which the corresponding terms are defined.  In the subspace spanned by the modes with $\ell_1, \ell_2\ge 1$, we find a single tower of negative modes associated with a particular mixing of $(\phi_I, \sigma_I)$, and all the remaining modes are positive. This structure is manifest in the following form of the action
\begin{align}
\left.\tilde S^{(2)}_{\rm dD}\right|_{\ell_1,\ell_2\ge 1} &=\sum_{I}\lambda_{I,1}\lambda_{I,2}\left(\lambda_I - \frac{2}{L^2}\right)v_I^2+\lambda_{I,1}\lambda_{I,2}\left(\lambda_I - \frac{2}{R^2}\right)(u_I^2+w_I^2+z_I^2)\nonumber\\
&+\lambda_{I,1}\lambda_{I,2}\left[\left(\frac{2Q}{R^2}w_I+\frac{a_I-\bar a_I}{\sqrt{2}}\right)^2+  \left(\frac{2Q}{R^2}u_I+\frac{b_I}{\sqrt{2}}\right)^2 +  \left(\frac{2Q}{R^2}z_I+\frac{\bar b_I}{\sqrt{2}}\right)^2 \right]\\
&+\frac{1}{2}(\lambda_{I, 1}a_I+\lambda_{I,2}\bar a_{I})^2+\frac{1}{2}\lambda_{I,1}^2 b_I^2+\frac{1}{2}\left(\lambda_{I,2}\bar b_I+\frac{4\sqrt{2}Q}{R^2}\sigma_I\right)^2+\frac{2(\lambda_I-\frac{2}{R^2})^2}{\lambda_I-2\Lambda}\sigma_I^2-\frac{\lambda_I -2\Lambda}{8}\tilde \phi_I^2~,\nonumber
\end{align}
where we have defined $\tilde\phi_I = \phi_I+\frac{4F^2}{\lambda_I-2\Lambda}\sigma_I$. The negative modes are associated with $\tilde\phi_I$ because $\lambda_I > \frac{2}{R^2}>2\Lambda$ in this case. 

Then we study modes with at least one of the angular momenta vanishing. For example, taking the $s$-wave limit of $\tilde S^{(2)}_{\rm dD}$ on the first sphere while keeping $\ell_2\ge 1$ yields
\begin{align}\label{reduction1}
\left.\tilde S^{(2)}_{\rm dD}\right|_{\ell_1=0, \ell_2\ge 1}&=\sum_{\ell_2\ge 1}\frac{1}{2}\lambda_{I,2}^2 \bar a_I^2+\frac{1}{2}\left(\lambda_{I,2}\bar b_I+\frac{4\sqrt{2}Q}{R^2}\sigma_I\right)^2+\frac{2(\lambda_{I,2}-\frac{2}{R^2})^2}{\lambda_{I,2}-2\Lambda}\sigma_I^2-\frac{\lambda_{I,2} -2\Lambda}{8}\tilde \phi_I^2.
\end{align}
In this subsector, all modes of $\tilde \phi$ are negative since  $\lambda_{I, 2}=\frac{\ell_2(\ell_2+1)}{R^2}>2\Lambda$ when $\ell_2\ge 1$. The situation is opposite for $\sigma$. All modes of $\sigma$ are positive  except for the $\ell_2=1$ mode because
 $\lambda_{I,2} = \frac{2}{R^2}$ if $\ell_2=1$. It indicates three zero modes of the action, corresponding to $\bar b_I = -2\sqrt{2}Q\sigma_I, \phi_I = -4\sigma_I$ for $I=(0,1)$ and all other modes vanishing. In terms of the original metric and U(1) gauge field variables, these modes are 
\begin{align}\label{ZM2}
\phi_{\mu\nu} = \begin{pmatrix} 0 & 0 \\ 0 & \sigma g_{\alpha'\beta'} \end{pmatrix}, \quad a_\mu = (0, \sqrt{2}Q \epsilon_{\alpha'\beta'}\nabla^{\beta'}\sigma) = (0,  \sqrt{2}R^2 F_{\alpha'\beta'}\nabla^{\beta'}\sigma)~,
\end{align}
where $\sigma$ is constant on the first sphere and has angular momentum 1 on the second sphere. According to the gauge transformation \eqref{g2}, the modes in \eqref{ZM2} can be generated by diffeomorphism along $\nabla_{\alpha'}(-R^2\sigma)$. In other words, the three zero modes of $\left.\tilde S^{(2)}_{\rm dD}\right|_{\ell_1=0}$ are diffeomorphisms along the conformal Killing vectors (CKV) of the second sphere.

Similarly, the $s$-wave reduction on the second sphere  gives
\begin{align}\label{reduction2}
\left.\tilde S^{(2)}_{\rm dD}\right|_{\ell_2=0} = \frac{1}{2}\lambda^2_{I, 1}(a_I^2+ b_I^2)+\frac{2(\lambda_{I,1}-\frac{2}{L^2})^2}{\lambda_{I,1}-2\Lambda}\sigma_I^2-\frac{\lambda_{I,1} -2\Lambda}{8}\tilde \phi_I^2~.
\end{align}
This action also has three zero modes, i.e. $\phi_I = 4\sigma_I$ for $I=(1, 0)$. The corresponding U(1) fluctuation vanishes and the corresponding metric fluctuation is supported on the first sphere $\phi_{\alpha\beta} = \sigma g_{\alpha\beta}$, where $\sigma$ carries angular momentum 1 on the first sphere and is a constant on the second sphere.
 These zero modes can be generated by diffeomorphisms along the CKVs of the first sphere. Unlike in the previous case, $\tilde\phi_I$ can have more positive modes depending on the magnitude of the magnetic charge. The positive modes satisfy                                                                                                                                                                                                                                                                                                                                                                       $\frac{\ell_1(\ell_1+1)}{L^2}<2\Lambda$, or equivalently 
 \begin{align}\label{Lbound}
 \ell_1(\ell_1+1)<1+\frac{1}{\sqrt{1-4\Lambda Q^2}}~.
 \end{align}
The $\ell_1=0$ and 1 modes are always positive. Each positive mode of $\tilde\phi_I$ corresponds to a negative mode of $\sigma$, except for the $\ell_1=1$ ones.

\,

\noindent{}\textbf{Summary}: The spectrum of $\tilde\phi$ consists of an infinite number of negative modes and a finite number of positive modes: $(0,0), (1,0), \cdots, (\ell_{1,*}, 0)$, where $\ell_{1,*}$ is the largest integer satisfying \eqref{Lbound}. The remaining fields have a positive spectrum except for 6 zero modes and a finite number of negative modes. The negative modes have angular momenta $(0,0), (2,0), \cdots, (\ell_{1,*}, 0)$, involving  mixing of $\sigma$ and $\phi$. Three of the zero modes carry angular momentum (1,0), also due to mixing of $\sigma$ and $\phi$. The other three zero modes are supported on the second sphere with angular momentum $(0,1)$. They are linear combinations of $\phi$, $\sigma$ and the U(1) fluctuations.
The 6 zero modes make it more subtle to count the phase of the full one-loop path integral.
We will present a careful treatment of these zero modes in the next section.

\section{Deformations of the de Donder gauge}
\label{defdeDon}
In the previous section, we find 6 zero modes of $S^{(2)}_{\rm dD}$. They are diffeomorphisms preserving the  de Donder gauge. However, they are just artifacts of this special gauge choice, and can be avoided by choosing a different gauge.
In this section, we consider a more general class of gauge-fixing terms 
\begin{align}
S_\xi= \int \xi\left[(\nabla\cdot\phi)_\mu - \frac{\xi+1}{4\xi}\nabla_\mu \phi\right]^2~.
\end{align}
The relative coefficient of $(\nabla\cdot\phi)_\mu$ and $\nabla_\mu\phi$ is chosen such that $S_\xi$ does not introduce mixing between the Weyl factor $\phi$ and other components of the metric fluctuation \cite{Shi:2025amq}. 
In particular, $\xi=1$ corresponds to the de Donder gauge. 
The gauge fixed action $S^{(2)} + S_\xi$ is by definition, equivalent to $S^{(2)}_{\rm dD} + \delta S_{\xi}$, where $\delta S_{\xi}=S_\xi- S_{\xi=1}$.
We are interested in the limit $|\xi-1|\ll 1$ because in this limit the non-zero modes of $S^{(2)}_{\rm dD}$ will not change signs under the deformation $\delta S_\xi$.  Therefore, most of the analysis in the previous section is still reliable and it suffices to study how the zero mode sector changes after adding $\delta S_\xi$
.

Half of the zero modes of $S^{(2)}_{\rm dD}$ correspond to CKVs on the first sphere. They are linear combinations  of $\phi_{1,0}$ and $\sigma_{1,0}$. Denote the subspace spanned by these 6 modes by $V_{1,0}$. The restriction of $S^{(2)}_{\rm dD} + \delta S_{\xi}$ to $V_{1,0}$ becomes \footnote{This restriction is well-defined because the only modes that carry angular momenta (1, 0) are $\{\phi_{1,0}, \sigma_{1,0}, a_{1, 0}, b_{1,0}\}$ and $\{a_{1, 0}, b_{1,0}\}$ do not mix with $\{\phi_{1,0}, \sigma_{1,0}\}$. In particular, note that there are no $\psi_{\alpha\beta}$ or $\psi_{\alpha'\beta'}$ modes with this angular momenta.}
\begin{align}\label{SV10}
    S_{V_{1, 0}}&=\frac{1}{8}F^2(\phi_{1,0}-4\sigma_{1, 0})^2 + \frac{2}{L^2}(\xi-1)\left(\sigma_{1,0}^2-\frac{\phi_{1,0}^2}{16\xi}\right)\nonumber\\
    &=\frac{1}{8}(\phi_{1, 0}, 4\sigma_{1, 0})\begin{pmatrix} F^2-\frac{\xi-1}{\xi L^2}& -F^2 \\ -F^2 & F^2+\frac{\xi-1}{L^2}\end{pmatrix}\begin{pmatrix} \phi_{1, 0}\\ 4\sigma_{1, 0}\end{pmatrix}~.
\end{align}
The  matrix in \eqref{SV10} has a positive eigenvalue $\lambda_1 = 2F^2+\CO\left((\xi-1)^2\right)$. It corresponds to a perturbation of the $\tilde\phi_{1,0}^2$ term in \eqref{reduction2}. The sign of the other eigenvalue depends on the magnitude of $Q$
\begin{align}
    \quad \lambda_2 = \left(1-\frac{1}{F^2 L^2}\right)\frac{(\xi-1)^2}{2L^2}+\CO\left((\xi-1)^3\right).
\end{align}
Using \eqref{LRsolu}, we find $F^2 L^2 =1/\sqrt{1-4\Lambda Q^2}-1$, and hence $\lambda_2<0$ when $0<Q<\frac{1}{4}\sqrt{3/\Lambda}$ and $\lambda_2>0$ when $\frac{1}{4}\sqrt{3/\Lambda}<Q<Q_{\rm max}$. In particular, at $Q_*=\frac{1}{4}\sqrt{3/\Lambda}$, $\lambda_2$ becomes zero, corresponding to $\phi_{1, 0} = 4\xi\sigma_{1, 0}$. In terms of the metric variable, these zero modes are 
\begin{align}
    a_\mu=0, \quad \phi_{\mu\nu} = \sigma_{1, 0}\begin{pmatrix}
        (\xi+1)g_{\alpha\beta} & 0 \\ 0 &  (\xi-1)g_{\alpha'\beta'}
    \end{pmatrix}
\end{align}
Both $S^{(2)}$ and $S_\xi$ vanish at these field configurations. The  component $(\xi+1)\sigma_{1, 0}g_{\alpha\beta}$ can  be removed  by  diffeomorphism in the first sphere. So, these zero modes are equivalent to $\phi_{\alpha'\beta'} = \sigma_{1,0} g_{\alpha'\beta'}$. They are physical zero modes. As pointed out in \cite{Chen:2025jqm}, the physical zero modes arise because the Nariai solution and the lukewarm solution coincide at the special charge $Q_*=\frac{1}{4}\sqrt{3/\Lambda}$, as shown in Figure \ref{skfin}.

The remaining zero modes of $S^{(2)}_{\rm dD}$ correspond to CKVs on the second sphere. They are linear combinations  of $\phi_{0,1}$, $\sigma_{0,1}$ and $\bar b_{0,1}$, which generate a 9-dimensional subspace $V_{0,1}$. The gauge fixed action in this subspace takes the form 
\begin{align}
    S_{V_{0, 1}} =\frac{2}{R^4} \left(\bar b_{0, 1}\!+\!2\sqrt{2}Q\sigma_{0,1}\right)^2 -\frac{(F^2 R^2\!+\!1)\xi-1}{8 R^2\xi}\left(\phi_{0,1}\!+\!\frac{4F^2 R^2\xi\,\sigma_{0,1}}{(F^2 R^2\!+\!1)\xi-1}\right)^2 +\frac{2(1\!+\!F^2 R^2)(\xi-1)^2\sigma_{0,1}^2}{R^2((F^2 R^2\!+\!1)\xi-1)}~.
\end{align}
The first term is the same as in \eqref{reduction1}, and the second term is a small perturbation of the $\tilde\phi_{0,1}^2$ term in \eqref{reduction1}.
The last term indicates that the three zero modes are lifted to positive modes for $\xi$ sufficiently close to 1.

Altogether, we find two different behaviors of the zero mode as we deform the de Donder gauge, depending on the magnitude of $Q$. When $0<Q<Q_*$, the 6 zero modes split into 3 negative modes and three positive modes. When  $Q_*<Q<Q_{\rm max}$, all zero modes are lifted to positive modes.
In the former case, each positive/negative mode in $\tilde\phi$ corresponds to a negative/positive mode in $\sigma$. Using the counting prescription proposed in \cite{Ivo:2025yek}, we find the total phase of the one-loop partition function to be 1. In particular, it is consistent with the $Q=0$ limit. In the latter case, as the $3$ negative modes in $\sigma$ become positive, the full one-loop partition function picks up a phase of $i^3$.

\section{Discussion}
\label{disc}

In this note, we analyzed the phase of the one-loop Euclidean path integral around charged Nariai solutions in 4 dimensions, confirming the result obtained from a dilaton gravity truncation in \cite{Chen:2025jqm}. Our result suggests that this is a consistent truncation for the calculation of the phase, at least in this specific context. 

The phase obtained at small values of the charge is consistent with the result for uncharged Nariai \cite{Ivo:2025yek, Shi:2025amq}. At a critical value of the charge, the phase transitions to $i^{3}$, which is curiously also the phase expected for a small charged black hole in de Sitter \cite{Maldacena:2024spf, Chen:2025jqm}. While we did not investigate ``lukewarm solutions" in the Shark Fin, the phase of these solutions is currently being worked on by a different group \cite{akhtar2025wip}. 

As a final remark, we now also briefly comment on the intriguing proposal of \cite{Chen:2025jqm} for an alternative way of obtaining state counting partition functions for an observer from the Euclidean path integral. Their proposed method differs from the one in \cite{Maldacena:2024spf} by a factor of $(-1)$, such that they obtain a positive ``state counting density of states" for an observer in de Sitter. This final phase for the density of states is very satisfactory. However, we stress that the way the answer is obtained raises an apparent puzzle at the level of negative mode counting, which was also briefly pointed out in their discussion \cite{Chen:2025jqm}. While we do not attempt to answer this puzzle here, we will flesh it out in detail in the hopes of attracting more attention to this issue.

To summarize the puzzle, we remind the readers that the sphere gravitational path integral with an observer has a phase of $i^{3}$ \cite{Maldacena:2024spf}. From our definitions of negative modes, this is equivalent to ``minus three" negative modes. However, as a ``last step", one would like to relate these path integrals to ``state counting" partition functions. Maldacena's \cite{Maldacena:2024spf} proposal on how to define a state counting partition function from the path integral effectively introduces an extra factor of $(-i)$ to the path integral, leaving overall two missing negative modes, e.g, $i^{2}=(-1)$ as the final phase. He then suggests that these two last modes could have something to do with the physics of the cosmological horizon. 

The authors of \cite{Chen:2025jqm} (CSTY) have a different proposal on how to perform this last step, which boils down to introducing a factor of $i$ instead of $(-i)$. Doing so, one obtains an overall phase of $i^{4}$, e.g, the phase of ``minus four negative modes". It of course holds that $i^{4}=1$, so one ends up with a positive density of states. However, this result seems to follow from the puzzling fact that ``minus four" negative modes have no overall phase. In other words, from the fact that the phase is only dependent on the number of negative modes $\text{mod } 4$. 

This is a different type of phase cancellation from the one that happens, for example, between the observer's phase and the dimension-dependent part of the phase from the sphere partition function in \cite{Maldacena:2024spf}. There, $D-1$ negative modes from the observer cancel with $D-1$ missing negative modes in the sphere partition function. This has the interpretation, as proposed by Maldacena \cite{Maldacena:2024spf}, that one can use the observer's degree of freedom to gauge fix the ``missing negative modes" in the sphere calculation. Therefore, one would, perhaps naively, expect that the mechanism for getting rid of the residual phase from the three missing negative modes in \cite{Maldacena:2024spf} would have an analogous geometric interpretation. That is, one would expect the mechanism to involve three extra negative modes cancelling these three missing ones.

This puzzle is also related to the question of whether one should judge the stability of the saddle based on its phase or instead on its number of negative modes. While one might naively expect the phase to be the only final observable, the number of negative modes can itself be important for determining whether a saddle is relevant. For example, the overall phase of a bounce instanton being $(-i)$ is famously linked to their tunneling interpretation\cite{Callan:1977pt}\footnote{The $\pm i$ is convention dependent.}. If the relevance of the tunneling interpretation were based on the phase alone, tunneling interpretations would be acceptable for saddles with $1,5,9$,... negative modes. However, as shown by Coleman \cite{Coleman:1987rm}, saddles with more than one negative mode have nothing to do with tunneling. This is, therefore, a context where the number of negative modes itself is relevant. 

An obvious wrinkle in the points that we raised is that in the discussion of negative modes in Euclidean gravity, the number of negative modes is formally infinite. When we discuss negative modes, we generally refer to a renormalized notion, defined after we subtract ``one negative mode per lattice point". It could then perhaps be that this renormalized notion of negative modes is itself ambiguous. Another possibility is that the fact that we have apparently four missing negative modes is an artifact of some specific way of evaluating the Euclidean path integral. Perhaps there is a way of relating the path integral with four missing negative modes to a better version of the path integral with none. In any case, since we think that the density of states for an observer in de Sitter is well defined, one would perhaps wonder if there is a version of the state counting partition function that has ``no negative modes" in the end. This is what we seem to get, for example, for partition functions in AdS that have a holographic interpretation.

More work towards a semiclassical understanding of de Sitter seems to be clearly necessary. Still, hopefully, the renewed interest in these issues is bringing us closer to a satisfactory understanding of de Sitter physics.

\subsection*{Acknowledgments}

We would like to thank Yiming Chen, Juan Maldacena, Douglas Stanford, Haifeng Tang, Jiuci Xu and Zhenbin Yang for very useful discussions. Z.S. is supported  by the U.S. Department of Energy
grant DE-SC0009988.

\bibliographystyle{utphys}

\bibliography{main.bib}

\end{document}